\documentstyle[amsfonts,aps]{revtex}

\begin{document}
\title{Quantum photon emission from a moving mirror in the nonperturbative 
regime}
\author{J. P. F. Mendon\c ca$^{1,3}$, P. A. Maia Neto$^2$ and F. I. 
Takakura$^1$}
\address{$^1$Departamento de F\'{\i}sica, ICE, Universidade Federal de Juiz de 
Fora, 
\\
36036-330 Juiz de Fora, Minas Gerais, Brazil\\
$^2$Instituto de F\'{\i}sica, Universidade Federal do Rio de Janeiro,\\
Caixa Postal 68528,\\
21945-970 Rio de Janeiro, Rio de Janeiro, Brazil\\
$^3$Department of Physics, Stanford University, Stanford, CA94305, USA}
\date{\today }
\maketitle

\begin{abstract}
We consider the coupling of the electromagnetic vacuum field with an
oscillating perfectly-reflecting  mirror in the nonrelativistic
approximation. As a consequence of the frequency modulation associated to
the motion of the mirror, low frequency photons are generated. We calculate
the photon emission rate by following a nonperturbative approach, in which
the coupling between the field sidebands is taken into account. We show that
the usual perturbation theory fails to account correctly for the
contribution of TM-polarized vacuum fluctuations that propagate along
directions nearly parallel to the plane surface of the mirror. 
 As a result of the modification of the field
eigenfunctions, the resonance frequency for photon emission is shifted from
its unperturbed value.
\end{abstract}

\begin{center}
PACS: 4250,  0365, 4210.
\end{center}

\section{Introduction}

When the boundaries of the quantum radiation field are set in motion, low
frequency photons may be excited out of the vacuum field state. Such quantum
effect is closely related to the radiation pressure exerted on the moving
boundaries by the vacuum field~\cite{ford}. Given its conceptual importance,
one has proposed an experiment to observe the effect with the aid of high-Q
microwave cavities~\cite{astrid}. Exact relativistic results are known for
one--dimensional models~\cite{1d}, but the analysis of realistic 3D
configurations usually demands some perturbation method. Several different
perturbation approaches~\cite{geral}--\cite{mundarain} have been
successfully employed to compute the radiation effect induced by the
nonrelativistic motion of a broad variety of 3D boundaries. In reference~%
\cite{las}, the photon emission rates induced by the oscillation of a single
perfectly-reflecting plane mirror were obtained by taking the long
wavelength approximation and assuming the motion induced corrections to be
small. More recently, this particular perturbation approach was also
employed in the problem of two plane mirrors forming a cavity~\cite
{mundarain}.

The connection between the nonrelativistic limit and the long wavelength
approximation may be understood by considering the example of a mirror
oscillating at frequency $\Omega _0$ with amplitude $\delta \!q_0.$ Then the
nonrelativistic limit translates into $\Omega _0\delta \!q_0\ll c,$ and
since the emitted photon frequencies satisfy the inequality $\omega \le
\Omega _0$~\cite{las}, they correspond to wavelengths $\lambda $ much larger
than $\delta \!q_0.$ In this long wavelength regime, the scattering of a
plane wave of frequency $\omega $ usually generates sidebands at frequencies 
$\omega -m\Omega _0$ with amplitudes of the order of $(\delta \!q_0/\lambda
)^{|m|}$, 
where $m$ is any integer. Hence the sidebands have amplitudes which
are increasingly small as the order $m$ increases. Moreover, the motion
induced correction of the elastic scattered component is usually of the
order of $(\delta\! q_0/\lambda)^2.$ Therefore, the photon emission effect
induced by a nonrelativistic motion may be analyzed by taking a first-order
expansion in $\delta\! q_0/\lambda,$ and then the entire motion effect is
contained in the sidebands at $\omega\pm\Omega_0.$ Accordingly, in this
approach the motion induced terms are treated as small perturbations which
may be computed directly from the field for a mirror at rest~\cite{las}.

However, when the incident wave is polarized so that the magnetic field is
perpendicular to the plane of incidence (TM polarization), the first
downshifted sideband ($m=-1$) may be intense if the parameters are such that
it corresponds to a grazing wave (peak amplitude value of the order of $%
\lambda /\delta \!q_0).$ Such enhancement effect, which is reminiscent of
the Wood's anomalies in the scattering by a diffraction grating at rest,
first explained by Rayleigh in the beginning of the century~\cite{Ray}, was
analyzed in reference~\cite{opt} from the point-of-view of classical wave
optics, always in the long wavelength approximation. It takes place over a
very narrow spectral interval, and may be interpreted in terms of resonances
associated to the plane symmetry ( in the context of diffraction gratings, a
discussion along these lines was proposed by Fano \cite{fano}). More
specifically, the resonances are related to surface waves (plasmons) which
are the field eigenfunctions for the unperturbed system (surface at rest).
In the perfect-reflection limit considered here, they are expelled from the
medium and then degenerate into grazing travelling waves. These unperturbed
eigenfunctions are TM polarized (so that the electric field is perpendicular
to the surface of the mirror), thus no resonance enhancement occurs for TE
scattering in the long wavelength approximation.

As for the motion induced quantum radiation effect, in order to compute the
emission rate taking these resonances into account, we follow in this
article a nonperturbative scheme, though still assuming a nonrelativistic
motion, which allows us to employ the long wavelength approximation. We
consider in detail the contribution of TM-polarized vacuum fluctuations
which propagate along directions nearly parallel to the surface of the
mirror. As we show below, the motion induced correction to TM-polarized
nearly grazing waves is not a small perturbation in this case, and the
dynamical coupling between different sidebands must be taken into account.
As a result the elastic field components are strongly modified, leading to
a shift of the resonance frequency for generation of photons. 

The paper is organized as follows. In section 2, we expand the scattered
field as a superposition of sidebands in order to derive a nonperturbative
solution of the boundary condition associated to an oscillating mirror,
which is then used in section 3 to compute the photon emission rate. Section
4 presents a discussion and some final remarks.

\section{Boundary conditions}

We consider a plane perfectly--reflecting mirror oscillating along the
direction perpendicular to its surface, which we take along the $x$ axis.
The mirror's position at time $t$ is given by the equation $x=\delta \!q(t).$
We use MKS units and take $\epsilon _0=1,$ $c=1.$ As usual we decompose the
electromagnetic field into components according to whether the electric
field is perpendicular (TE) or parallel (TM) to the plane of incidence. As a
consequence of the plane symmetry, the two polarizations are uncoupled in
the scattering by the moving mirror, and accordingly we have two separate
problems to solve. In the case of TE polarization, the perturbation approach
is always valid in the long wavelength regime (as in the problem of
reflection by a shallow diffraction grating), so that we only consider in
this paper the generation of TM photons out of the vacuum radiation field.
We define the vector potential $\mbox{\boldmath$\cal A$}$ as follows: 
\begin{eqnarray}
{\bf E}^{{\rm \scriptstyle(TM)}} &=&\nabla \times \mbox{\boldmath$\cal A$},
\label{ete} \\
{\bf B}^{{\rm \scriptstyle(TM)}} &=&\partial _t\mbox{\boldmath$\cal A$}.
\label{etm}
\end{eqnarray}
Moreover, it satisfies the Gauge condition 
\begin{equation}
\nabla \cdot \mbox{\boldmath$\cal A$}=0.  \label{gauge}
\end{equation}
The key motivation for defining the vector potential as in eqs.~(\ref{ete}%
)--(\ref{gauge}) is the fact that $\mbox{\boldmath$\cal A$}\cdot \hat{x}=0,$
which entails that $\mbox{\boldmath$\cal A$}$ does not change under a
Lorentz boost. For the Lorentz frame in which the mirror is instantaneously
at rest the boundary condition for $\mbox{\boldmath$\cal A$}$ simply states
that its normal derivative vanishes at the mirror's position. Then,
translating back to the laboratory frame, such condition yields (see~\cite
{mundarain} for details): 
\begin{equation}
\lbrack \partial _x+\delta \!\dot{q}\partial _t]\mbox{\boldmath$\cal A$}%
(\delta \!q(t),{\bf r_{\parallel }},t)=0,  \label{bc1}
\end{equation}
where ${\bf r_{\parallel }}=(y,z).$

The field is written as 
\begin{equation}
{\mbox{\boldmath$\cal A$}}={\mbox{\boldmath$\cal A$}}_{{\rm in}}+\delta \!{%
\mbox{\boldmath$\cal A$}}_{{\rm ret}}  \label{def1}
\end{equation}
where $\delta \!{\mbox{\boldmath$\cal A$}}_{{\rm ret}}$ is the retarded
field that represents the effect of the motion of the mirror, whereas ${%
\mbox{\boldmath$\cal A$}}_{{\rm in}},$ representing the input field ($%
t\rightarrow -\infty $), is itself a sum of incident and reflected waves,
and hence satisfies the boundary conditions of a mirror at rest. We take
periodic boundary conditions on the plane of the mirror over a square of
surface $S,$ which is to be identified with the surface of the (very large)
mirror later. Therefore, the normal mode decomposition of ${%
\mbox{\boldmath$\cal A$}}_{{\rm in}}$ is given by 
\begin{equation}
{\mbox{\boldmath$\cal A$}}_{{\rm in}}(x,{\bf r_{\parallel }}%
,t)=\int_0^\infty {\frac{dk_x}{2\pi }}\sum_n\sqrt{\frac{2\hbar }{k_nS}}\cos
(k_xx)e^{i{{\bf k}_{\parallel }}_n\cdot {\bf r_{\parallel }}}e^{-ik_nt}{a_{%
{\rm in}}}_n(k_x){\hat \epsilon }_n+{\rm H.c.},  \label{modes1}
\end{equation}
where H. c. means the Hermitian conjugate, ${{\bf k}_{\parallel }}_n=(n_y{%
\hat{y}}+n_z{\hat{z}})2\pi /\sqrt{S},$ with $n\equiv (n_y,n_z)$ denoting a
pair of integer numbers, ${\hat{{\bf \epsilon }}}_n={\hat{{\bf x}}}\times {%
\hat{{\bf k}}_{\parallel }}{}_n$ is the polarization vector, and $k_n=\sqrt{%
k_x^2+{k_{\parallel }}_n^2}$ is the frequency of a given normal mode. The
input bosonic operators satisfy the commutation relations 
\begin{equation}
\lbrack {a_{{\rm in}}}_n(k_x),{a_{{\rm in}}}_{n^{\prime }}(k_x^{\prime })]=0
\label{comm1}
\end{equation}
and 
\begin{equation}
\lbrack {a_{{\rm in}}}_n(k_x),{a_{{\rm in}}}_{n^{\prime }}(k_x^{\prime
})^{\dagger }]=2\pi \delta (k_x-k_x^{\prime })\delta _{n,n^{\prime }}.
\label{comm2}
\end{equation}

We assume that the oscillatory motion of the mirror is nonrelativistic: $%
\delta \!\dot{q}\ll 1.$ As discussed in the introduction, this assumption
allows us to take the long wavelength approximation in order to solve for $%
\delta \!{\mbox{\boldmath$\cal A$}}_{{\rm ret}}$ in terms of ${%
\mbox{\boldmath$\cal A$}}_{{\rm in}}.$ Thus, we expand the fields around $%
x=0 $ keeping terms up to order of $\delta \!q/\lambda ,$ and find, from
eqs.~(\ref{bc1}) and (\ref{def1}), 
\begin{equation}
\left( \partial _x+\delta \!q\partial _x^2+\delta \!\dot{q}\partial
_t\right) \delta \!{\mbox{\boldmath$\cal A$}}_{{\rm ret}}(0,{\bf %
r_{\parallel }},t)=-\left( \delta \!q\partial _x^2+\delta \!\dot{q}\partial
_t\right) {\mbox{\boldmath$\cal A$}}_{{\rm in}}(0,{\bf r_{\parallel }},t)
\label{bc2}
\end{equation}
In the perturbative regime considered in reference~\cite{las}, $\delta \!{%
\mbox{\boldmath$\cal A$}}_{{\rm ret}}$ is already of first--order in $\delta
\!q/\lambda ,$ and then only the first term in the lhs of eq.~(\ref{bc2})
contributes to the required order of $\delta \!q/\lambda .$ In this paper,
we consider a regime, involving propagation near the grazing direction,
where $\delta \!{\mbox{\boldmath$\cal
A$}}_{{\rm ret}}$ may be of a higher order over a very short spectral range.
In fact, eq.~(\ref{bc2}) itself provides a hint that it is not possible to
neglect the last two terms in its lhs when considering waves propagating
along the grazing direction, since $\partial _x\delta \!{\mbox{\boldmath$%
\cal A$}}_{{\rm ret}}$ vanishes in this case. These terms provide a coupling
between the motion induced sidebands, which is to be analyzed later in this
section.

It is convenient to define a Fourier representation that takes advantage of
the plane symmetry and the corresponding conservation of the parallel
component of the wavevector, ${{\bf k}_{\parallel }}_n.$ Hence we define the
Fourier components of a vector field ${\bf F}({\bf r},t)$ as 
\begin{equation}
{\bf F}_n[x,\omega ]={\frac 1S}\int dt\int_Sd^2r_{\parallel }\ e^{i\omega
t}\ e^{-i{{\bf k}_{\parallel }}_n\cdot {\bf r_{\parallel }}}{\bf F}(x,{\bf r}%
_{\parallel },t),  \label{fou}
\end{equation}
and the wave equation reads 
\begin{equation}
\partial _x^2{\bf F}_n[x,\omega ]=({k_{\parallel }}_n^2-\omega ^2){\bf F}%
_n[x,\omega ].  \label{waveeq}
\end{equation}
The normal mode expansion of eq.~(\ref{modes1}) is written as 
\begin{equation}
{\mbox{\boldmath$\cal A$}}_{{\rm in}}{}_n[x,\omega ]=\Theta ({\cal K}(\omega
)^2)\sqrt{\frac{2\hbar \left| \omega \right| }{{\cal K}(\omega )^2S}}\cos (%
{\cal K}(\omega )x)\left[ \Theta (\omega )a_{{\rm in}_n}({\cal K}(\omega
))-\Theta (-\omega )a_{{\rm in}_{-n}}^{\dagger }(-{\cal K}(\omega ))\right] {%
\hat{{\bf \epsilon }}}_n,  \label{modes2}
\end{equation}
where 
\begin{equation}
{\cal K}(\omega )=\lim\limits_{\varepsilon \rightarrow 0^+}[(\omega
+i\varepsilon )^2-k_{{\parallel }_n}^2]^{1/2}  \label{cut}
\end{equation}
is a function in the complex plane of $\omega $ with a branch cut along the
interval in the real axis between $-k_{\parallel }$ and $k_{\parallel }.$ It
represents the $x$--component of the reflected wave associated to the
parameters $\omega $ and ${{\bf k}_{\parallel }}_n.$ The factor $\Theta (%
{\cal K}(\omega )^2)$ in eq.~(\ref{modes2}) (where $\Theta $ denotes the
Heaviside step function) is a consequence of excluding evanescent waves from
the mode expansion of eq.~(\ref{modes1}), whereas the factor $\Theta (\omega
)$ establishes the connection between annihilation (creation) operators and
positive (negative) frequencies, which is crucial for the understanding of
the photon emission process discussed in this paper.

The mirror oscillates around the position $x=0:$ 
\begin{equation}
\delta \!q(t)=\delta \!q_0\cos \Omega _0t,  \label{sin}
\end{equation}
with $\Omega _0\delta \!q_0\ll 1,$ leading to the generation of sidebands at
frequencies 
\[
\omega _m=\omega -m\Omega _0. 
\]
We take the Fourier transform of eq.~(\ref{bc2}) as defined by eq.~(\ref{fou}%
) and use eq.~(\ref{waveeq}) to find (from now on we omit explicit reference
to ${\bf k}_{{\parallel }_n}$): 
\begin{equation}
\partial _x\delta \!{\mbox{\boldmath$\cal A$}}_{{\rm ret}}[0,\omega ]+\frac 1%
2\delta \!q_0\sum_{j=-1,1}\left[ \left( k_{\parallel }{}_n^2-\omega \omega
_j\right) \left( \delta \!{\mbox{\boldmath$\cal A$}}_{{\rm ret}}[0,\omega
_j]+{\mbox{\boldmath$\cal A$}}_{{\rm in}}[0,\omega _j]\right) \right] =0.
\label{bc3}
\end{equation}
In order to solve eq.~(\ref{bc3}), the retarded field is written as a
superposition of all sidebands generated from the input field. In the
half-space corresponding to $x>0,$ we have 
\begin{equation}
\delta \!{\mbox{\boldmath$\cal A$}}_{{\rm ret}}[x,\omega ]=\exp \left( i%
{\cal K}(\omega )x\right) \sum\limits_{m=-\infty }^\infty g_m(\omega _{-m}){%
\mbox{\boldmath$\cal A$}}_{{\rm in}}[0,\omega _{-m}].  \label{flo}
\end{equation}
According to eqs.~(\ref{cut}) and (\ref{flo}) (and to the particular choice
of the branch cut on the complex plane associated with the former), $\delta
\!{\mbox{\boldmath$\cal A$}}_{{\rm ret}}[x,\omega ]$ either corresponds to a
wave propagating from the mirror into the half-space $x>0$ (when $|\omega
|>k_{\\}$) or to an evanescent wave decaying along the positive $x$
direction, since ${\cal K}$ is of the form ${\cal K}(\omega )=i|{\cal K}%
(\omega )|$ when $|\omega |<k_{\\}.$

We replace eq.~(\ref{flo}) into eq.~(\ref{bc3}) and use eq.~(\ref{cut}) to
find, after changing the summation index: 
\begin{equation}
\sum\limits_{m=-\infty }^\infty \left\{ i{\cal K}(\omega )g_m(\omega _{-m})-{%
\frac{\delta \!q_0}2}\sum_{j=-1,1}([{\cal K}(\omega _j)]^2+j\omega _j\Omega
_0)(g_{m+j}(\omega _{-m})+\delta _{m,-j})\right\} {\mbox{\boldmath$\cal A$}}%
_{{\rm in}}[0,\omega _{-m}]=0.  \label{inter}
\end{equation}
Since eq.~(\ref{inter}) applies to any input field, the expression within
curly brackets must vanish for all values of $m$ and arbitrary values of $%
\omega .$ When considering a given value of $m$ in eq.~(\ref{inter}), we
replace $\omega $ by $\omega _m=\omega -m\omega _0$ in order to have the $g$
functions evaluated at the same frequency $\omega .$ We
obtain an infinite system of coupled linear equations for the functions $%
g_m(\omega )$ of the form 
\begin{equation}
{\Bbb M}\text{{\bf g}}={\bf Y},  \label{sistema}
\end{equation}
where ${\Bbb M}$ is the symmetric tridiagonal infinite matrix given by 
${\Bbb M}_{m,m}=i{\cal K(\omega }_m),$
\[
{\Bbb M}_{m+1,m}={\Bbb M}_{m,m+1}={\frac{\delta \!q_0}2H(}{\cal \omega }_m), 
\]
and with  ${\Bbb M}_{m,m^{\prime }}=0$
otherwise. 
We have introduced the auxiliary function 
\[
H(\omega )=\omega \Omega _0-{\cal K(\omega )}^2, 
\]
and
${\bf Y}$ and ${\bf g}$ are column vectors, with 
\[
{\bf Y}_m=-{\frac{\delta \!q_0}2H(\omega })\delta _{m,1}-{\frac{\delta \!q_0}%
2H(}\omega _{-1})\delta _{m,-1}, 
\]
while the elements of ${\bf g}$ are the functions $g_m(\omega ),$ {\bf \ 
}the integer index $m$ running from $-\infty $ to $\infty .$ We may
recover the perturbation results by neglecting the nondiagonal elements of
the matrix ${\Bbb M}$, thus resulting in uncoupled equations whose solution
is 
\begin{equation}
g_1(\omega )\approx \frac i2\,\delta \!q_0{\frac{H(\omega )}{{\cal K}(\omega
_1)}},g_{-1}(\omega )\approx \frac i2\,\delta \!q_0{\frac{H(\omega _{-1})}{%
{\cal K}(\omega _{-1})},}  \label{g1lin}
\end{equation}
and with all other functions, including $g_0(\omega ),$ of higher order of $%
\omega \delta \!q_0.$ Note that these results are in agreement with the
analogous expressions of reference \cite{opt}, and could be more easily
obtained directly from eq.~(\ref{bc2}) by assuming $\delta \!{%
\mbox{\boldmath$\cal A$}}_{{\rm ret}}$ to be a small perturbation.

We seek a solution that interpolates this perturbative regime, considered in
detail in reference~\cite{las}, with the regime corresponding to the
scattering of vacuum fluctuations propagating along the grazing direction,
always in the long wavelength approximation. Thus, we assume that the $x$
component of the input field is small: $|{\cal K}(\omega )|\ll \omega
^2\delta \!q_0.$ In this case, it is no longer possible to neglect the
nondiagonal terms in the row corresponding to $m=0$, so that we cannot
neglect the coupling between $g_0,$ $g_1$ and $g_{-1}.$ We then solve eq.~(%
\ref{sistema}) with $m$ running from $-1$ to $1$~\cite{notaad}: 
\begin{equation}
g_1(\omega )=\frac i2\frac{{\cal K}(\omega )}{{\cal K}(\omega _1)}\,\frac{%
\delta \!q_0H(\omega )}{{\cal K}(\omega )+\left( \frac{\delta \!q_0}2\right)
^2\left( \frac{H(\omega )^2}{{\cal K}(\omega _1)}{\ +}\label{res1}\frac{%
H(\omega _{-1})^2}{{\cal K}(\omega _{-1})}\right) },  \label{g1}
\end{equation}
\begin{equation}
g_{-1}(\omega )=\frac{{\cal K}(\omega _1)H(\omega _{-1})}{{\cal K}(\omega
_{-1})H(\omega )}g_1(\omega )  \label{g-1}
\end{equation}
and 
\begin{equation}
g_0(\omega )+1=-2i\frac{{\cal K}(\omega _1)}{\delta \!q_0H(\omega )}%
g_1(\omega ).  \label{g0}
\end{equation}
The perturbative results as given by eq.~(\ref{g1lin}) are recovered from
eqs.~(\ref{g1})-(\ref{g0}) when $|{\cal K}(\omega )/\omega |\gg (\omega
\delta \!q_0)^2.$ On the other hand, since $g_0$ is of the order of one when 
$|{\cal K}(\omega )/\omega |\le (\omega \delta \!q_0)^2,$ the oscillation of
the mirror generates a strong modification of the elastic field components
when considering propagation close to a grazing direction.
In fact, eq.~(\ref{g0}) yields $g_0(\omega )=-1$ when ${\cal K}(\omega )=0.$
Thus, in this limit, we have ${\mbox{\boldmath$\cal A$}}={%
\mbox{\boldmath$\cal A$}}_{{\rm in}}+\delta \!{\mbox{\boldmath$\cal A$}}_{%
{\rm ret}}$ $=0$, showing that in this case the retarded field $\delta \!{%
\mbox{\boldmath$\cal
A$}}_{{\rm ret}}$ is of course not a small perturbation~\cite{nota}, and
that the TM grazing waves, which are the field eigenfunctions in the case of
a perfectly-reflecting mirror at rest, are no longer allowed in the motional
case. This effect is at the origin of the  resonance frequency
shift for quantum photon generation 
out of the vacuum field state, to be discussed in the next two sections.

\section{Photon emission rate}

In this section, we use the nonperturbative results for the retarded field
found in section 2 in order to consider the effect of photon generation
induced by the motion of the mirror. As discussed in connection with eq.~(%
\ref{modes2}), changing from positive to negative frequencies yields a
coupling between annihilation and creation photon operators, which is
responsible for the radiation effect considered in this paper.

In order to analyze such coupling, we start by writing the total field, as
given by eq.~(\ref{def1}), in terms of the advanced solution $\delta \!{%
\mbox{\boldmath$\cal A$}}_{{\rm adv}}$ of the boundary condition. In this
case, the homogeneous component of the solution is interpreted as an output
field, which represents the limit $t\rightarrow \infty :$ 
\begin{equation}
{\mbox{\boldmath$\cal A$}}_{{\rm in}}[x,\omega ]+\delta \!{%
\mbox{\boldmath$\cal A$}}_{{\rm ret}}[x,\omega ]={\mbox{\boldmath$\cal A$}}_{%
{\rm out}}[x,\omega ]+\delta \!{\mbox{\boldmath$\cal A$}}_{{\rm adv}%
}[x,\omega ].  \label{in-out1}
\end{equation}
The output field ${\mbox{\boldmath$\cal A$}}_{{\rm out}}$ may be expanded in
terms of output bosonic operators (also satisfying the commutation relations
of eqs.~(\ref{comm1}) and (\ref{comm2})) and normal modes exactly as in
eqs.~(\ref{modes1}) and (\ref{modes2}), because it also satisfies the
homogeneous boundary condition associated to a mirror at rest. Moreover, the
advanced solution may be expanded as in eq.~(\ref{flo}): 
\begin{equation}
\delta \!{\mbox{\boldmath$\cal A$}}_{{\rm adv}}[x,\omega ]=\exp \left( -i%
{\cal K}(\omega )^{*}x\right) \sum\limits_{m=-\infty }^{+\infty }f_m(\omega
_{-m}){\mbox{\boldmath$\cal A$}}_{{\rm out}}[0,\omega _{-m}],  \label{flo2}
\end{equation}
where the coefficients $f_m$ are given by $f_m=g_m^{*}.$

Eq.~(\ref{in-out1}) establishes a connection between input and output
bosonic operators. In what concerns the dependence on the spatial coordinate 
$x,$ note that the evanescent components contained in both $\delta \!{%
\mbox{\boldmath$\cal
A$}}_{{\rm ret}}$ and $\delta \!{\mbox{\boldmath$\cal A$}}_{{\rm adv}}$
cancel each other in eq.~(\ref{in-out1}) and do not contribute to the photon
emission effect, as expected since the radiation field is associated to
travelling waves only. Then, we take $|\omega |\geq k_{\parallel }{}_n$ and
assume ${\cal K}(\omega )$ to be real in what follows. As a consequence, the
fields may be written entirely in terms of the two linearly independent
functions $\exp (i{\cal K}(\omega )x)$ and $\exp (-i{\cal K}(\omega )x)$.
Since eq.~(\ref{in-out1}) applies to any value of $x,$ the coefficient
multiplying $\exp (i{\cal K}(\omega )x)$ must vanish, yielding, with the aid
of eq.~(\ref{modes2}): 
\begin{eqnarray}
\sqrt{\frac{\hbar |\omega |}{2{\cal K}(\omega )^2S}}\Bigl\{\Theta (\omega
)\left[ a_{{\rm out}_n}({\cal K}(\omega ))-a_{{\rm in}_n}({\cal K}(\omega
))\right] &&  \nonumber \\
-\Theta (-\omega )\left[ a_{{\rm out}_{-n}}(-{\cal K}(\omega ))-a_{{\rm in}%
_{-n}}(-{\cal K}(\omega ))\right] ^{\dagger }\Bigr\}{\hat{{\bf \epsilon }}}%
_n &=&\delta \!{\mbox{\boldmath$\cal A$}}_{{\rm ret}}[0,\omega ].
\label{bog1}
\end{eqnarray}
If we take $\omega <-k_{\parallel }{}_n$ we single out the creation
operators in the lhs of eq.~(\ref{bog1}). However, the retarded field in its
rhs evaluated at $\omega $ may also contain annihilation operators, since $%
\omega $ may correspond to a downshifted sideband associated to an initially
positive frequency $\omega +\Omega _0.$ Such mixture between creation and
annihilation operators entails that photons are created from vacuum by means
of the frequency modulation associated to the motion of the mirror. In
Appendix A, we start from eq.~(\ref{bog1}) and compute the average value of
the output number operator associated to given values of ${\bf k}_{{%
\parallel }_n}$ and $k_x$ over the input vacuum field state $|0\,{\rm in}%
\rangle $ (we use the shorthand $\langle ...\rangle $ to denote such
average): 
\begin{equation}
\left\langle a_{{\rm out}_n}^{\dagger }(k_x)a_{{\rm out}_n}(k_x)\right%
\rangle =\frac{4k_x^2}k\Theta (\Omega _0-k-{\ k}_{{\parallel }_n}){\frac{%
\left| g_1(\Omega _0-k)\right| ^2}{{\cal K}(\Omega _0-k)}}\Delta \!t,
\label{av3}
\end{equation}
where $\Delta \!t$ is a coarse-grained time interval, and $k=\sqrt{k_x^2+{k}_{{\parallel }%
_n}^2}$ is the photon frequency. Eq.~(\ref{av3}) explicitly relates the
photon creation effect to frequency downshifting, which is dominated by $g_1$
in the long wavelength approximation (an equivalent description in terms of
frequency upshifting is also possible and leads to the same final results).
The step function in its r.-h.-s. expresses the requirement that the input
vacuum fluctuations at frequency $\Omega _0-k$ correspond to traveling
waves. We consider a direction of emission forming an angle $\theta $ with
the $x$ axis, so that $k_{\parallel _n}=k\sin \theta .$ We analyze the
behaviour of the photon emission rate at fixed frequency $k$ and propagation
angle $\theta $ as we tune the mechanical frequency $\Omega _0.$ It is
convenient to define the dimensionless variable 
\[
\Delta \equiv \frac{\Omega _0}k-1-\sin \theta . 
\]
According to eq.~(\ref{av3}), no photons are generated when $\Delta <0.$ At
such low values of the mechanical frequency, there are no travelling wave
vacuum field modes at frequency $\Omega _0-k$ that match the requirement
associated to the conservation of the parallel component of the wavevector, $%
{\bf k}_{\parallel _n}.$ As a consequence, regardless of the choice of $%
\theta ,$ the mechanical frequency must be higher than the analyzed photon
frequency $k.$ As could be expected, high frequency field modes are
unaffected by the motion of the mirror (quasistatic limit). This basic fact
is essential to the long wavelength approximation employed in the paper.

At $\Delta=0,$ the vacuum fluctuations that contribute to the photon
emission effect correspond to grazing waves. More generally, when $\Delta
\ll 1$ the wavevectors of the relevant vacuum fluctuations have their $x$
component given by 
\begin{equation}
{\cal K}(\Omega_0 - k)= k \sqrt{2\sin\theta\,\Delta\,}[1+{\cal O}(\Delta)].
\label{deltapeq}
\end{equation}
In the perturbative approximation, the function $g_1$ remains finite (see
eq.~(\ref{g1lin})) and then the photon number as given by eq.~(\ref{av3})
diverges as $\Delta\rightarrow 0$ since ${\cal K}(\Omega_0-k)$ vanishes in
this limit. Ref.~\cite{las} provides a simple interpretation of the
divergence at $\Delta=0.$ In the perturbative theory, photons are emitted in
pairs. When the mechanical frequency is set to $\Delta=0,$ the observed
photon is such that its `twin' propagates along a grazing direction~\cite
{note}. TM polarized grazing waves are eigenfunctions for the
perfectly-reflecting mirror at rest --- in fact they represent the limit of
the surface plasmons of a conducting surface as we take the
perfectly--reflecting limit. Thus, $\Delta=0$ is the resonance frequency for
the field given the boundary condition of a perfectly-reflecting plane
mirror at rest, for in this case the external modulation perfectly matches
the conditions for the creation of grazing photons.

On the other hand, according to the nonperturbative result of eq.~(\ref{g1}%
), the coefficient $g_1(\Omega_0-k)$ is proportional to ${\cal K}%
(\Omega_0-k) $ as $\Delta\rightarrow 0.$ Therefore, instead of the
divergence at $\Delta=0 $ predicted by the perturbative theory, the photon
emission rate vanishes in this limit. In Appendix B, we compute the 
angular distribution 
rate for $\Delta\ge 0$ from eq.~(\ref{av3}): 
\begin{equation}
\frac{d^2R}{dkd{\it \Omega }}=\frac S{2\pi ^3}\frac{k^2\cos ^2\theta }{\sqrt{%
\Delta (\Delta +2\sin \theta )}}\left| g_1\left[ \omega =k(\Delta +\sin
\theta )\right] \right| ^2.  \label{final}
\end{equation}
It follows from eqs.~(\ref{g1}), (\ref{deltapeq}) and (\ref{final}) that the
photon emission rate goes as $\sqrt{\Delta}$ in the limit $\Delta\ll
(k\delta\!q_0)^4.$ On the other hand, according to the discussion following
eqs.~(\ref{g1})--(\ref{g0}), the perturbative regime is recovered when the
vacuum fluctuations propagate along a direction not close to a grazing
direction: ${\cal K}(\Omega_0 - k)/k\gg (k\delta\!q_0)^2,$ which from eq.~(%
\ref{deltapeq}) translates into $\Delta\gg (k\delta\!q_0)^4.$ In this limit,
we recover the results 
of Ref.~\cite{las}.

Since $g_1$ is dimensionless, it depends on $k$ only through $\Delta $ and $%
k\delta \!q_0.$ Thus, apart from the trivial $k^2$ dependence in eq.~(\ref
{final}), we may replace $\delta \!q_0,$ $\Omega _0$ and $k$ by the two
dimensionless parameters $k\delta \!q_0$ and $\Delta .$ In figure 1, we plot
the photon emission rate as given by eq.~(\ref{final}) as function of $%
\Delta $ for $\theta =78^0$ and $k\delta \!q_0=0.03.$ In the main figure, we
also plot the result from perturbation theory (dotted line), which is
obtained from eqs.~(\ref{g1lin}) and (\ref{final}). As discussed above,
whereas the latter diverges at $\Delta =0$, the exact emission rate (solid
line) vanishes in this limit. However, the exact result also displays a
singularity, which is shifted to a higher value 
$\Delta_s.$ Thus, the resonance frequency for photon emission is
shifted from its unperturbed value $\Omega_0=k(1+\sin\theta)$ by an
amount $$\delta\!\Omega=k\Delta_s.$$
 The exact value of $\Delta _s$  is best displayed in
the insert of figure 1, where we compare the analytical result as given by
replacing eq.~(\ref{g1}) into eq.~(\ref{final}) (dashed line) with the
result obtained from the numerical evaluation (solid line) of the coupled
equations (\ref{sistema}) for $g_m$ with $m$ running from $-3$ to $3$ (we
have checked that truncations at higher dimensions had not changed the
results within machine accuracy). The insert shows that the 
exact frequency shift
is slightly smaller than the analytical value for the
particular values of $\theta$ and $k\delta\!q_0$ considered
in fig.~1. On the other hand, 
outside the
neighbourhood of $\Delta _s$ the analytical result works fairly well, and
in fact for the scale employed in the main plot of 
fig.~1 the two methods provide
indistinguishable curves. 

From eq.~(\ref{g1}), we may derive an analytical approximation for 
$\Delta_s$ up to lower order of $k\delta\!q_0:$ 
\begin{equation}
\Delta_s={\frac{(k\delta\!q_0)^4}{32}} \sin^3\theta(1+\sin\theta)^4 \left({%
\frac{1}{\cos\theta}}- {\frac{1}{\sqrt{3\sin^2\theta+4\sin\theta+1}}}%
\right)^2  \label{deltas}
\end{equation}
For $\theta=78^0,$ and $k \delta\!q_0 = 0.03,$ eq.~(\ref{deltas}) yields $%
\Delta_s=7.187\times 10^{-6},$ which is in agreement with the plot of the
emission rate as given by the analytical result (dashed line in the insert
of fig. 1). In figure 2, we plot $\Delta_s$ as given by eq.~(\ref{deltas})
as a function of the observation angle $\theta.$ The shift vanishes at $%
\theta=0$ and increases by several orders of magnitude as $\theta$ increases
from zero to $90^0.$ In the next section, we discuss such behaviour starting
from the analysis of the diagram representing the input vacuum fluctuations
that contribute to the radiation emission effect.

\section{Discussion}

In the previous section, we discussed the behaviour of the photon emission
rate for fixed values of the photon frequency and direction of emission, as
we tune the mechanical frequency $\Omega _0.$ In this section, we consider
the complementary situation, in which $\Omega _0$ is fixed, in order to
understand why the resonance frequency is shifted from its 
unperturbed value.
For the
sake of clarity, we analyze the photon emission process as an effect of
frequency upshifting (rather than downshifting) in this section.

Each TM-polarized photon have given values of 
frequency
$\omega ,$ parallel component of wavevector $k_{\parallel},$ and an
 azimuthal angle
defining the direction of ${\bf k}_{\parallel}.$
 The photons generated from a mechanical modulations at
frequency $\Omega _0$ correspond to points in the ABC triangle (dark grey)
in the $\omega \times k_{\parallel }$ plane of figure~3, since they obey the
inequalities $\omega +k_{\parallel }\le \Omega _0$ and $\omega \ge
k_{\parallel }\ge 0.$ Normal modes with $\omega <k_{\parallel }$ do not
correspond to travelling waves and hence
 are not relevant here.
They correspond to the light grey region in figure 3.

In the $\omega\times k_{\parallel}$ plane the 
frequency upshifting induced by the
motion corresponds to a horizontal shift (since $k_{\parallel}$ is
conserved) by an amount of $\Omega_0.$ Hence the vacuum fluctuations that
contribute to the emission effect correspond to the A'B'C' triangle in
figure 3. Vacuum fluctuations associated to the B'C' line propagate along
grazing directions. They give rise to photons associated to points along the
BC line, all of them corresponding to $\Delta=0.$ In this paper, we have
shown that in the neighbourhood of B'C' the perturbative approach fails to
provide the amplitudes of the
sidebands. Whereas the perturbative theory predicts a divergent emission
rate all along the BC line, we have found that the emission rate vanishes in
this case, and that the singularity is displaced into the interior
neighbourhood of the BC border.

This effect 
is a consequence of the motion induced modification 
of the field
eigenfunctions. For a metallic
surface {\it at rest},
the eigenfunctions correspond to  surface plasmons. In
the perfect-reflecting limit, they are expelled from the interior of the
medium and degenerate into TM polarized grazing waves. In the framework of
first-order perturbation theory, the dynamical modification of the field
eigenfunctions is neglected, 
and then the 
resonant enhancement takes place when considering vacuum fluctuations 
along the B'C' line, which give rise to photons in the field modes 
corresponding to $\Delta=0.$ However, this resonant regime is
inconsistent with the
underlying assumption that the motion effect is a small perturbation. In
this paper, we have employed a nonperturbative approach, and found the
correction of the elastic field component, which is represented by the
coefficient $g_0$ in eq.~(\ref{flo}), to be very important in the case of
nearly grazing waves (see eq.~(\ref{g0}) and the discussion 
that follows).  Thus,  TM polarized
grazing waves are no longer acceptable solutions in the motional case. Since
 the field eigenfunctions are
strongly modified,  the resonant region
 is displaced away from the B'C'
line in figure 3 into the interior of the A'B'C' region. As a consequence,
the critical region in the positive-frequency part of the diagram is
displaced from the $\Delta=0$ line (BC line), leading to a shift of the
resonance frequency.

Vacuum fluctuations associated to the A'B' border give rise to photons
propagating along a grazing direction (AB line). In this case, the
perturbative approach also fails to provide the correct photon emission rate
(see remark~\cite{nota}). The two critical regions overlap at point B, which
corresponds to grazing photons at the subharmonic frequency, generated from
grazing vacuum fluctuations. As we approach B along the BC line, the width
of the nonperturbative region increases, as well as the frequency shift $%
\Delta _s.$ In fact, that corresponds to the limit $\theta \rightarrow 90^o$
discussed in connection with eq.~(\ref{deltas}) (see fig.~2). Note, however,
that eq.~(\ref{deltas}) was derived from eq.~(\ref{g1}) by assuming $\Delta
_s\ll 1.$ Moreover, eq.~(\ref{g1}) itself was obtained from the coupled
equations (\ref{sistema}) by neglecting higher order sidebands, which turns
out not to be a good approximation in this limit. As shown in the insert of
fig.~1, the exact shift is smaller than the value given by eq.~(\ref{deltas}%
), and the discrepancy increases as $\theta \rightarrow 90^o.$

The ABC triangle is closed from below by the AC border. The latter
corresponds to photons that propagate along the normal direction. Since they
are generated from vacuum fluctuations that also propagate along the normal
direction (A'C' line), the nonperturbative theory provides no relevant
correction in this case. As expected, according to fig.~2 the frequency
shift $\Delta_s$ vanishes as we approach the neighbourhood of point C ($%
\theta\rightarrow 0$).

In this paper, we have analyzed in detail the resonance effect associated to
the excitation of surface plasmons from the vacuum field state when a mirror
oscillates in free space.  Due to the assumption of perfect reflectiveness,
the surface plasmons degenerate into TM-polarized grazing waves, and the
resonance appears as a singularity in the photon emission rate. The
nonperturbative approach developed in this paper allowed us to calculate the
shift of the resonance frequency induced by the motion of the mirror, which
is a consequence of the dynamical modification of the field eigenfunctions.

The authors thank Programa Especial de Visitante (PREVI-UFJF),  Conselho Nacional de Desenvolvimento
Cient\'{\i}fico e Tecnol\'{o}gico (CNPq) for partial support (J. P. R. F. 
M. and P. A. M. N.) and Programa de N\'ucleos de Excel\^encia (PRONEX), 
grant no.  4.1.96.08880.00 -- 7035-1 (P. A. M. N.).

\appendix

\section{Average values of output number operators}

In this appendix, we derive eq.~(\ref{av3}) for the average value of the
output number operator. We replace the retarded field in eq.~(\ref{bog1}) by
its expansion in terms of sidebands as given by eq.~(\ref{flo}) and use the
normal mode decomposition as written in eq.~(\ref{modes2}). Then, as
mentioned in sec. 3, we pick up a frequency $\omega<-{k}_{{\parallel }_n}$
to find 
\begin{eqnarray}
\left\langle a_{{\rm out}_{-n}}^{\dagger }(-{\cal K}(\omega ))a_{{\rm out}%
_{-n}}(-{\cal K}(\omega ))\right\rangle &=&4\Bigm|{\frac{\omega _{-1}}\omega 
}\Bigm| \left[ {\frac{{\cal K}(\omega )}{{\cal K}(\omega _{-1})}}\right]
^2\Theta (\omega _{-1}-{k}_{{\parallel }_n})\left| g_1(\omega _{-1})\right|
^2  \label{av1} \\
& &\times \left\langle a_{{\rm in}_n}({\cal K}(\omega _{-1}))a_{{\rm in}%
_n}^{\dagger }({\cal K}(\omega _{-1})) \right\rangle.  \nonumber
\end{eqnarray}
Note that it is also possible to compute the photon emission rate by picking
up a positive $\omega $ in eq.~(\ref{bog1}), but in this case the dominant
contribution comes from the function $g_{-1},$ which describes frequency
upshifting. Of course this method leads to the same final result for the
average output number operator.

In order to calculate the vacuum correlation function in the rhs of eq.~(\ref
{av1}), we introduce a coarse--grained time scale $%
\Delta \!t,$ which satisfies $\Omega_0 \Delta\!t \gg 1,$ by means
of a frequency distribution sharply peaked around the mechanical frequency $%
\Omega _0:$ 
\begin{equation}
h(\Omega )=\frac 1\pi \frac{1/\Delta \!t}{(\Omega -\Omega _0)^2+(\frac 1{%
\Delta \!t})^2}.  \label{h}
\end{equation}
The distribution $h(\Omega )$ is normalized so as to yield 
\[
\int d\Omega \,h(\Omega )=1. 
\]
It represents the spectral profile associated to the mirror's motion: $%
\delta \!q[\Omega]=\delta \!q_0[h(\Omega )+h(-\Omega )]/2.$ Therefore,
rather than the sinusoidal motion of eq.~(\ref{sin}), we actually have an
exponentially damped sinusoidal motion (where $\Delta \!t$ is the damping
time). In the frequency domain, it amounts to replace the delta function by
the distribution $h.$ Since $h(\Omega )$ is very sharply peaked around $%
\Omega _0,$ however, we may replace $\Omega $ by $\Omega _0$ everywhere (and
thus keep all the development of sec. 2), except when computing the vacuum
correlation function in eq.~(\ref{av1}). For instance, the representation of
eq.~(\ref{flo}) is replaced by 
\[
\delta \!{\mbox{\boldmath$\cal A$}}_{{\rm ret}}[x,\omega ]=\exp \left( i%
{\cal K}(\omega )x\right) \sum\limits_{m=-\infty }^\infty g_m(\omega
+m\Omega _0)\int h(\Omega ){\mbox{\boldmath$\cal A$}}_{{\rm in}}[0,\omega
+m\Omega ]d\Omega , 
\]
and moreover $\Omega $ is replaced by $\Omega_0$ everywhere in ${%
\mbox{\boldmath$\cal A$}}_{{\rm in}}[0,\omega +m\Omega ]$ except in the
argument of the bosonic input operators. The finite mechanical linewidth is
taken into account only when computing the vacuum correlation function, for
which we take an average over the distribution $h:$ 
\begin{equation}
\left\langle a_{{\rm in}_{-n}}({\cal K}_{-1})a_{{\rm in}_n}^{\dagger }({\cal %
K}_{-1})\right\rangle =\int d\Omega ^{\prime }\int d\Omega ^{^{\prime \prime
}}h(\Omega ^{^{\prime }})h(\Omega ^{^{\prime \prime }})\left\langle 0\,{\rm %
in}\right| a_{{\rm in}_n}({\cal K}(\omega +\Omega ^{^{\prime }}))a_{{\rm in}%
_n}({\cal K}(\omega +\Omega ^{^{\prime \prime }}))^{\dagger }\left| 0\,{\rm %
in}\right\rangle .  \label{vac2}
\end{equation}
The unaveraged correlation function is easily calculated from the
commutation relations given by eq.~(\ref{comm2}): 
\begin{equation}
\left\langle 0\,{\rm in}\right| a_{{\rm in}_n}({\cal K})a_{{\rm in}_n}({\cal %
K}^{\prime })^{\dagger }\left| 0\,{\rm in}\right\rangle =2\pi \delta ({\cal K%
}-{\cal K}^{\prime }).  \label{vac3}
\end{equation}
Replacing eq.~(\ref{vac3}) into eq.~(\ref{vac2}) and using the auxiliary
result 
\[
\int d\Omega \,h^2(\Omega )=\frac{\Delta \!t}{2\pi } 
\]
yields 
\begin{equation}
\left\langle a_{{\rm in}_n}({\cal K}(\omega _{-1}))a_{{\rm in}_n}^{\dagger }(%
{\cal K}(\omega _{-1}))\right\rangle =\frac{\left| {\cal K}(\omega
_{-1})\right| }{\left| \omega _{-1}\right| }\Delta\!t.  \label{vac}
\end{equation}

When inserting eq.~(\ref{vac}) into (\ref{av1}), it is convenient to employ
the variables $k_x$ and ${\bf k}_{{\parallel }_n}$ defining a given
wavevector ${\bf k}$ and the corresponding bosonic operator ${a_{{\rm in}}}%
_n(k_x)$ [see eq.~(\ref{modes1})], rather than the Fourier variable $\omega
. $ Thus, we take $k_x=-{\cal K}(\omega ),$ and then $k=\sqrt{k_x^2+{\ k}_{{%
\parallel }_n}^2}=-\omega .$ Note that since $\omega <-{\ k}_{{\parallel }%
_n} $ eq.~(\ref{cut}) yields $k_x>0$ as it should. The final result is given
by eq.~(\ref{av3}).

\section{Connection between photon emission rates and number operators}

In this appendix, we establish the connection between the averaged number
operator, as given by eq.~(\ref{av3}) and the photon emission rate. The
field normal mode decomposition as given by eq.~(\ref{modes1}) is such that
the lhs of eq.~(\ref{av3}) represents, when multiplied by $dk_x/(2\pi ), $
the average number of (TM polarized) photons with wavevectors whose $x$
component is between $k_x$ and $k_x+dk_x$ and whose parallel component is $%
{\bf k}_{{\parallel }_n}.$ Since the number of allowed values of ${\bf k}_{{%
\parallel }_n}$ per unit area in the two--dimensional $yz$ reciprocal space
is $S/(2\pi )^2,$ the average number of photons with frequency between $k$
and $k+dk$ is 
\begin{equation}
d^3N=\left\langle a_{{\rm out}_n}^{\dagger }(k_x)a_{{\rm out}%
_n}(k_x)\right\rangle {\frac{dk_x}{2\pi }}d^2{\bf k}_{{\parallel }_n}{\frac S%
{(2\pi )^2}}.  \label{rate1}
\end{equation}
According to eqs.~(\ref{av3}) and (\ref{rate1}), $d^3N$ is proportional to
the coarse-grained time interval $\Delta \!t,$ which allows us to define the
photon production {\it rate} $d^2R=d^3N/\Delta \!t.$ Writing the volume
element in spherical coordinates, 
\begin{equation}
d^3{\bf k}=dk_xd^2{\bf k}_{{\parallel }_n}=k^2dkd{\it \Omega }{\sc ,}
\label{vol}
\end{equation}
where ${\it \Omega }$ denotes the solid angle, we obtain from eq.~(\ref
{rate1}) the connection between the angular distribution rate of photon
emission (photon emission rate per frequency and solid angle intervals) and
the averaged number operator: 
\begin{equation}
\frac{d^2R}{dkd{\it \Omega }}=\frac{Sk^2}{(2\pi )^3\Delta \!t}\left\langle
a_{{\rm out}_n}^{\dagger }(k_x)a_{{\rm out}_n}(k_x)\right\rangle .
\label{rate2}
\end{equation}
By replacing eq.~(\ref{av3}) into (\ref{rate2}), we obtain the result for
the photon emission rate as given by eq.~(\ref{final}).

\newpage 

{\bf Figure Captions} \bigskip\bigskip

Figure 1. Photon emission rate as function of the normalized mechanical
frequency $\Delta $ (see text), with $\theta =78^o$ and $k \delta\!q_0 =0.03.
$  Main plot: as calculated from the perturbation theory (dotted line) and
from the nonperturbative approach presented in this paper (solid line).
Insert plot: numerical results (solid line) and analytical approximation
where high order sidebands are neglected (dashed line). 
Logarithmic scale is employed in the vertical axis in the insert. 
Note that for the
scale employed in the main plot, it is not possible to distinguish numerical
and analytical curves.

Figure 2. Frequency shift $\Delta_s$ as calculated from the analytical
nonperturbative theory as a function of observation angle $\theta.$ When $%
\theta$ varies from $0$ to $90^o,$ $\Delta_s$ increases by several orders of
magnitude.

Figure 3. $\omega \times k_{\parallel }$ plane representing the field modes.
For a given mechanical frequency $\Omega _0,$ the region bounded by the
triangle ABC (dark grey) represents the  modes that may be excited.
The light grey region represents evanescent field modes (not relevant in the
present discussion). In this diagram, the photon creation process is
associated to frequency upshifting. This is represented by a horizontal
displacement (since $k_{\parallel }$ is conserved) of length $\Omega _0.$
Accordingly, photons in field modes along the BC line ($\Delta =0$) are
created from vacuum fluctuations that propagate along grazing directions
(B'C' line).


\begin{references}
\bibitem{notaad}  We have also compared eqs. (\ref{g1})-(\ref{g0}) with a
numerical solution of eq.~(\ref{sistema}) taking into account the coupling
with higher order $g$ functions. As discussed in the section 3, we have
found good agreement as far as the photon emission rate is concerned.
\bibitem{nota}  A similar effect takes place when the parameters are such
that the first downshifted sideband propagates along the grazing direction.
In this regime, considered in detail in ref. \cite{opt}, we have $\left| 
{\Bbb M}_{1,1}\right| =\left| {\cal K}(\omega _1)\right| \ll \omega ^2\delta
\!q_0,$ and then the best approximation is to consider the coupled equations
for $g_0$, $g_1$ and $g_2.$
\bibitem{note}  In fact, the twin photon is generated in the same field mode
associated to the vacuum fluctuation giving rise to the emission at $k$ and $%
\theta ,$ except for the fact that it propagates in the opposite sense.
\end{references}
\end{document}